\documentclass
[aps,prb,twocolumn,floatfix,english,showpacs,10pt]{revtex4-2}%
\usepackage[utf8]{inputenc}
\usepackage{amsmath}
\usepackage{physics}
\usepackage{braket}
\usepackage{amssymb}
\usepackage{graphicx}
\usepackage{xcolor}
\usepackage[most]{tcolorbox}
\usepackage[unicode=true, breaklinks=false, pdfborder={0 0 1}, backref=false, colorlinks=true, linkcolor=blue, urlcolor=blue, citecolor=blue]{hyperref}
\setcitestyle{numbers,square}

\begin{document}
\title{Gating ferromagnetic resonance of magnetic insulators by superconductors via modulating electric-field radiation}

\author{Xi-Han Zhou}

\affiliation{School of Physics, Huazhong University of Science and Technology, Wuhan 430074, China}

\author{Tao Yu}
\email{taoyuphy@hust.edu.cn}
\affiliation{School of Physics, Huazhong University of Science and Technology, Wuhan 430074, China}

\date{\today }

\begin{abstract}
We predict that ferromagnetic resonance in \textit{insulating} magnetic film with inplane magnetization radiates electric fields polarized along the magnetization with opposite amplitudes at two sides of the magnetic insulator, which can be modulated strongly by adjacent superconductors. With a single superconductor adjacent to the magnetic insulator this radiated electric field is totally reflected with a $\pi$-phase shift, which thereby vanishes at the superconductor side and causes no influence on the ferromagnetic resonance. When the magnetic insulator is sandwiched by two superconductors, this reflection becomes back and forth, so the electric field exists at both superconductors that drives the Meissner supercurrent, which in turn shifts efficiently the ferromagnetic resonance. We predict an ultrastrong coupling between magnons in the yttrium iron garnet and Cooper-pair supercurrent in NbN with a frequency shift achieving tens of percent of the bare ferromagnetic resonance.  
\end{abstract}

\maketitle

\section{Introduction}

``Magnonics" exploits magnetic excitations, i.e., spin waves or their quanta, magnons, as potential information carriers for spin transport in insulators with low-energy consumption~\cite{Lenk,Chumak,Grundler,Demidov,Brataas,Barman,Yu_chirality,Cornelissen,Zou,Wang}.
Interaction between magnons and Cooper-pair supercurrent in heterostructures composed of magnets and superconductors may modulate the transport of spin information~\cite{spintronics_1,superconductor_gating_theory,superconductor_gating_exp,silaev,Bobkova,FI/s_system,Wei_Han,Linder,Belzig,similar_theory}, strongly enhance the magnon-photon interaction~\cite{Janssonn_1,Janssonn_2,spintronics_2,strong_coupling_3,Silaev_ultrastrong_coupling,ultrastrong_in_press,Volkov}, and lead to the emergence of triplet Cooper pairing~\cite{Bergeret,study_1,study_2,Banerjee,Balatsky}, which may bring unprecedented functionalities in spintronics~\cite{Bergeret,study_1,study_2}, quantum information~\cite{Q_information,Q_information_1,Q_information_2,Q_information_3,Q_information_6,Q_information_4,Q_information_5}, and topological quantum computation~\cite{topological}. In this heterostructure, the hybridized quantum states and distribution of macroscopic electromagnetic fields govern its properties.
For example, the ``ultrastrong coupling"~\cite{strong_coupling} with the coupling strength close to the ferromagnetic resonance (FMR) frequency unveils the importance of the dipolar interaction in the  superconductor(S)$|$metallic ferromagnet(F)$|$superconductor(S) heterostructure~\cite{spintronics_2,strong_coupling_3,Volkov}, where the photon mode with a large mode density is localized in the nano-scale between two superconductors~\cite{Swihart}.

The importance of the dipolar interaction  also manifests in the superconductor gating effect on magnons~\cite{superconductor_gate,superconductor_gate_1,superconductor_gate_2,superconductor_gate_3,similar_theory,superconductor_gating_theory,superconductor_gating_exp}, in which the frequency of magnons with finite wave number~\cite{spin_wave,spin_wave_1,spin_wave_2,spin_wave_3} can be shifted  up to tens of GHz, as recently predicted~\cite{superconductor_gating_theory,similar_theory} and observed~\cite{superconductor_gating_exp} in the superconductor(S)$|$ferromagnet  
insulator(FI) heterostructure. 
The stray electric field of magnons drives the supercurrent in the adjacent superconductor which in turn generates the Oersted magnetic field that affects the low-frequency magnetization dynamics. This gating effect favors the spin diode~\cite{Yu_chirality,Yu_chirality_1} and magnon trap~\cite{Chumak_trap,magnon_trap,magnon_trap_1} in proper gating configurations. The FMR frequency in this S$|$FI bilayer is not affected, however.

\begin{figure}[htp]
    \centering 
    \includegraphics[width=\linewidth]{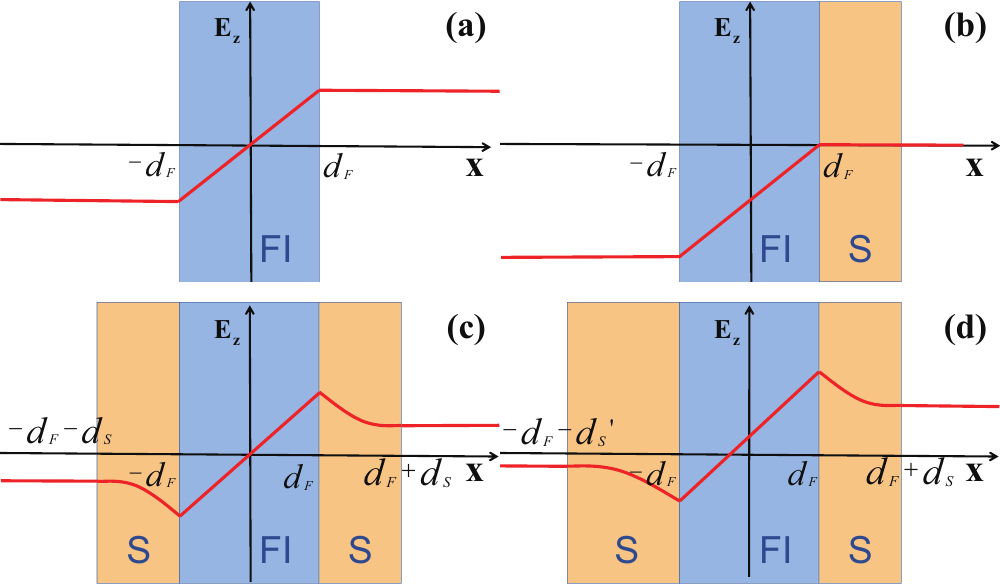}
\caption{Snapshots of magnetization-radiated electric fields in different heterostructure configurations. The electric field changes linearly across the thickness of the ferromagnetic insulating film. (a) The electric-field amplitude is opposite at two sides of the thin magnetic insulator.  (b) When fabricating a superconductor thin film on a  ferromagnetic insulator, the electric field is suppressed to vanish at the superconductor side but enhanced at the other side of the magnet. 
When the magnet is sandwiched by two
superconductors, the electric field exists but differs at both sides in both symmetric (c) and asymmetric (d) configurations.}
\label{electric_field}
\end{figure}

On the other hand, the FMR of the metallic ferromagnet sandwiched by two superconductors was shifted up to 50~mT in the resonant field when the thickness of the two superconductor layers is larger than the London's penetration depth, as observed in several recent experiments~\cite{CPL_exp,PRA_exp,experiment}. Above the superconducting transition temperature, the FMR frequency recovers to the  Kittel mode~\cite{kittel_mode}, which may be exploited to realize the magnetic logic gate through a phase transition in the superconductor.  This phenomenon may be related to the frequency splitting induced by spin-triplet superconducting state~\cite{CPL_exp},  Meissner screening~\cite{experiment}, and giant demagnetization effects~\cite{Gient_de,silaev}.  It appears that this modulation could be absent for the FMR in the ferromagnetic insulators~\cite{CPL_exp,experiment,Gient_de,silaev}, however, which has not been reported in the experiments yet~\cite{S_FI_S_junction_1,S_FI_S_junction_2,S_FI_S_junction_3}. Silaev predicted recently ultrastrong coupling between magnons and microwave photons in a magnetic insulator when sandwiched by two superconductors of infinite thickness, where the radiation of the electric field out of the heterostructure is completely suppressed~\cite{Silaev_ultrastrong_coupling}. The experiment~\cite{CPL_exp} showed that inserting a thin insulator layer in the heterostructures composed of a metallic ferromagnet sandwiched by two superconductors completely suppresses the shift of FMR. This raises the issue of whether the FMR can be gated or not in magnetic insulators by adjacent superconductors in proper configurations.

In this work, we study this issue by going beyond the quasi-static approximation for magnetostatic modes~\cite{Rezende} and demonstrate that although the stray magnetic field of Kittel magnon with uniform magnetization precession is vanishingly small outside of the in-plane magnetized ferromagnetic insulating film, the radiated electric field is significant with opposite amplitudes at two sides of the magnetic film and polarization parallel to the magnetization direction. This distribution of the radiated electric field is sensitive to the adjacent superconductors due to the total reflection, as illustrated in Fig.~\ref{electric_field} for snapshots of the distribution of electric fields in different heterostructure configurations.  
The electric field is opposite at two sides of a single thin ferromagnetic insulator [Fig.~\ref{electric_field}(a)]; contra-intuitively, in the S$|$FI bilayer this electric field is suppressed to vanish at the superconductor side  [Fig.~\ref{electric_field}(b)], when the superconductor thickness is larger than a nanometer; nevertheless, when sandwiched by two superconductors, the electric field is neither shifted to vanish nor screened completely, as plotted in Figs.~\ref{electric_field}(c) and (d) for symmetric and asymmetric configurations. These features are well understood by our mechanism of modulated reflection of magnetization-induced electric fields by superconductors, which predicts the absence of FMR shift in ferromagnetic insulator$|$superconductor heterostructure and the ultrastrong modulation of FMR, shifted up to tens of percent of the bare frequency when the ferromagnetic insulator is sandwiched by two thin superconductors.

This paper is organized as follows. We address the model and general formalism in Sec.~\ref{model_formalism}. In Sec.~\ref{single_ferromagnet}, \ref{bilayer}, and \ref{trilayer}, we analyze the distribution of the electric fields from FMR of a single ferromagnetic insulator, S$|$FI bilayer, and S$|$FI$|$S heterostructure, respectively, and address the ultrastrong interaction between the FMR and supercurrent. We conclude and discuss in Sec.~\ref{conclusion}.

\section{Model and general formalism}
\label{model_formalism}

\begin{figure}[h]
    \centering
\includegraphics[width=0.95\linewidth]{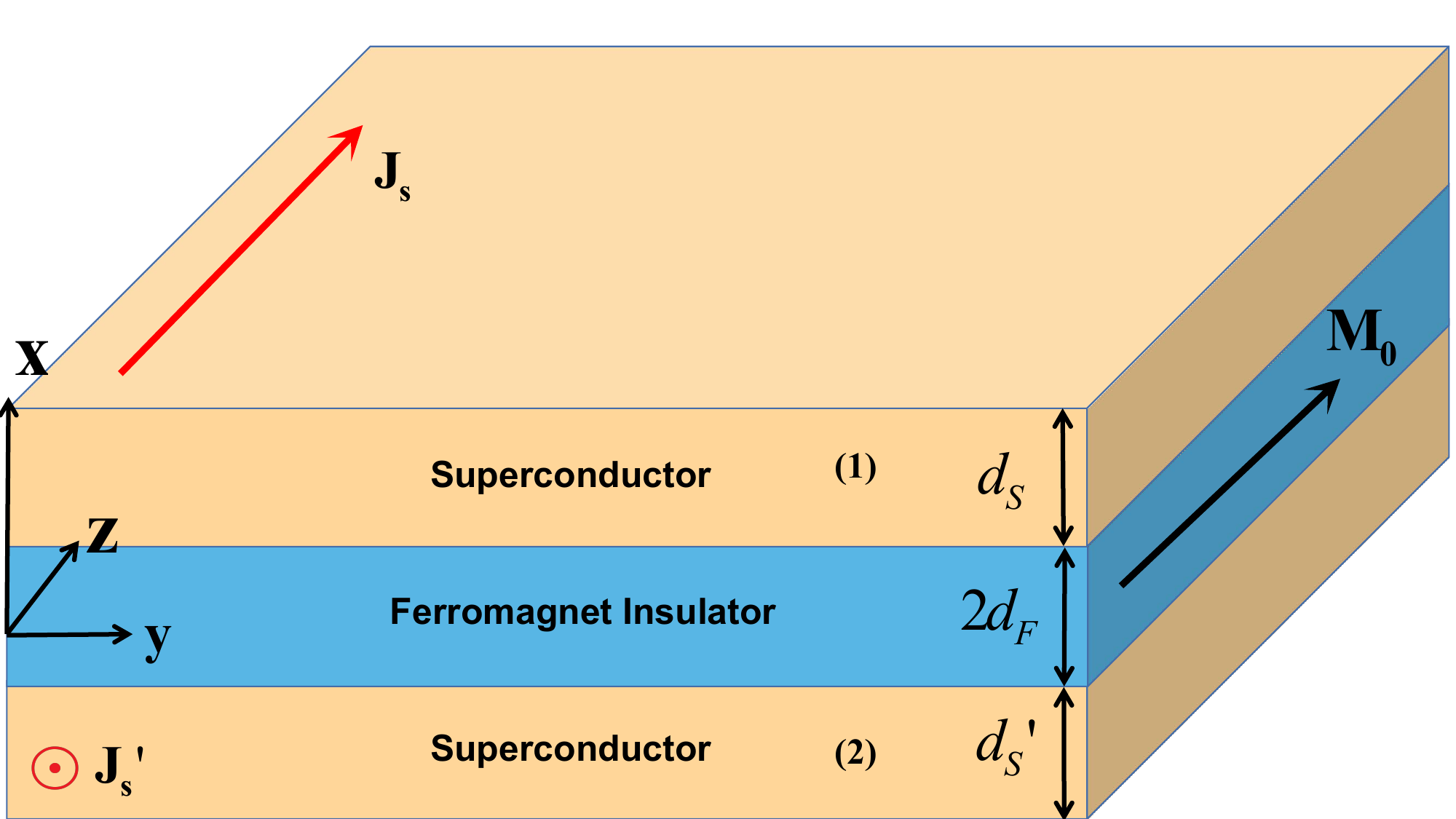}    \caption{S(1)$|$FI$|$S(2) heterostructure. The thickness of superconductors above and beneath the thin ferromagnetic insulator of thickness $2d_F$ is $d_S$ and $d_S'$, respectively. The driven supercurrents ${\bf J}_s$ and ${\bf J}_s'$ by FMR flow oppositely along the magnetization direction.}
    \label{model2}
\end{figure}

We consider a heterostructure composed of a ferromagnetic insulating film of thickness $2d_F\sim O(100~{\rm nm})$ with inplane magnetization sandwiched by two thin superconductor layers with thickness $d_S\lesssim \lambda$ and $d_S'\lesssim \lambda$, respectively, as illustrated in Fig.~\ref{model2}. Here $\lambda\sim O(100~{\rm nm})$ is London's penetration depth of conventional superconductors. In the ferromagnetic insulators, the dynamics of magnetization ${\bf M}=M_x\hat{\bf x}+M_y\hat{\bf y}+M_0\hat{\bf z}$, where $M_0$ is the saturated magnetization, is phenomenologically governed by the Landau-Lifshitz-Gilbert (LLG) equation~\cite{LLequation}
\begin{align}
    \partial {\bf M}/\partial t=-\mu_0\gamma {\bf M}\times{\bf H}+\alpha_G({\bf M}/M_0)\times\partial{\bf M}/\partial t,
    \label{Full_LLG_Equation}
\end{align}
where $\mu_0$ is the vacuum permeability, $-\gamma$ is the electron gyromagnetic ratio, and $\alpha_G$ is the damping coefficient of the magnetic insulator. The magnetization precesses around the effective magnetic field ${\bf H}= {\bf H}_{\text{app}}+{\bf H}_r$ that contains the external static field ${\bf H}_{\text{app}}=H_0 \hat{\bf z}$ and the radiated dynamic field ${\bf H}_r$ generated by the ``magnetic dipole radiation"~\cite{Jackson,superconductor_gating_theory}.  The energy flow out of the magnetic insulator then causes the radiation damping since the radiated magnetic field out of phase of the magnetization can exert a damping-like torque on the magnetization. The exchange interaction plays no role in the FMR since the gradient of ${\bf M}$ vanishes for the uniform precession.

The oscillating magnetic induction ${\bf B}=\mu_0({\bf M}+{\bf H})$  governs the radiation of electric fields inside and outside the ferromagnetic insulator according to~\cite{Jackson}
\begin{align}
\nabla\times {\bf E}=-\frac{\partial {\bf B}}{\partial t},&&
    \nabla\times {\bf H}={\bf J}_s
    +\varepsilon_0\frac{\partial {\bf E}}{\partial t},
\label{Maxwell_electric_field}
\end{align}
where $\varepsilon_0$ is the vacuum permittivity. When coupled with superconductors, this electric field drives the supercurrent ${\bf J}_s$ via London's equation~\cite{London_equation}  
\begin{align}
     \dfrac{\partial {\bf J}_s}{\partial t}= \dfrac{1}{\mu_0\lambda^2}{\bf E},&&
     \nabla\times {\bf J}_s=-\dfrac{1}{\mu_0\lambda^2}{\bf B}.
     \label{London}
\end{align} 
Here London's penetration depth at different temperatures $T<T_c$ follows the relation~\cite{London_equation}
\begin{align}
    \lambda(T)=\lambda_0\left(1-\left(\frac{T}{T_c}\right)^4\right)^{-1/2},
    \label{penetration_depth}
\end{align}
where $\lambda_0$ is London's penetration depth at zero temperature.

The boundary condition describes the fields at the interfaces~\cite{Jackson}. For the magnetic induction and field, ${\bf B}_\perp$ and ${\bf H}_\parallel$ are continuous at the boundaries. 
 Since there is no surface current or charge accumulation, the electric field $\bf E$ is continuous at interfaces.

At low frequencies and with near fields, the quasi-static approximation is usually applied~\cite{Jackson}, in which situation the radiation damping should be negligibly small. This is proved according to the calculation of radiation damping in Sec.~\ref{single_layer_full_solution}.  It is then sufficient to express the radiated magnetic field ${\bf H}_r$ as the summation of the dipolar field ${\bf H}_d$ and the Oersted field ${\bf H}_s$ from the superconductor~\cite{LLequation,Rezende}. The dipolar field
\begin{align}
\nonumber
{\bf H}_{d,\beta}({\bf M})&=\dfrac{1}{4\pi}\partial_\beta\sum_{\alpha}\partial_\alpha\int d{\bf r'}\dfrac{M_\alpha(\bf r')}{|\bf r-r'|}\\
&=\dfrac{1}{4\pi}\partial_\beta\int d{\bf r'}\dfrac{-\rho_m({\bf r'})}{|\bf r-r'|}
\label{H(M)}
\end{align}
is governed by Coulomb's law in terms of the magnetic charge $\rho_m=-\nabla\cdot{\bf M}$.

With the quasi-static approximation, $\nabla\times {\bf B}=\mu_0{\bf J}_s$ in superconductors. Taking the curl of Eq.~(\ref{Maxwell_electric_field}) and substituting Eq.~(\ref{London}) into it, the electric field inside the superconductor obeys
    \begin{align}
    \nabla^2 {\bf E}-{\bf E}/\lambda^2=0.
    \label{electric_field_in_s}
\end{align}
On the other hand, taking the curl of  $\nabla\times {\bf B}=\mu_0
{\bf J}_s$ and combining with Eq.~(\ref{London}), the magnetic induction inside the superconductor obeys 
$\nabla^2 {\bf B}-{\bf B}/\lambda^2=0$. 
The driven supercurrent then affects the magnetization dynamics. From Eq.~(\ref{London}), the electric field drives supercurrent inside the superconductor, which then generates the vector potential. With the uniform magnetization precession, the system is translational invariant in the $y$-$z$ plane, so the supercurrent only depends on $x$ and as it for the vector potential~\cite{Jackson} 
\begin{align}
  {\bf A}(x)=\dfrac{\mu_0}{4\pi}\int d{\bf r'}\dfrac{{{\bf J}_s}(x')}{|{\bf r}-{\bf r}'|}.
  \label{vector_potential}
\end{align}
Accordingly, the Oersted magnetic field 
\begin{align}
    {\bf H}_s=(1/\mu_0)\nabla\times {\bf A}
\end{align}
only contains the $y$-component ${H}_y=-\partial_xA_z(x)/\mu_0$, which drives the magnetization.

\section{Single thin ferromagnetic insulator}

\label{single_ferromagnet}

We start with a single insulating ferromagnetic film to address the significant radiated electric fields from the uniform magnetization precession. For a single ferromagnetic insulator of thickness $2d_F$  biased by a static magnetic field ${\bf H}_{\rm app}=H_0 \hat{\bf z}$,  the magnetization $\bf M$ for the FMR is uniform inside the ferromagnetic layer by the constant demagnetization factor $N_{xx}=-1$. Since the magnetic film is sufficiently thin, we stick to the uniform precession throughout this work.  The opposite magnetic charges at the two surfaces of the film generate opposite magnetic field outside, which results in vanished stray magnetic field ${\bf H}_d=0$ outside the ferromagnetic layer, as also calculated from Eq.~(\ref{H(M)});  inside the ferromagnet, ${\bf H}_d=\{-M_x,0,0\}$ and ${\bf B}=\{0,\mu_0 M_y,\mu_0(H_0+M_0)\}$, in which only the $y$-component of $\bf B$  oscillates with frequency $\omega$ that can radiate the electric field.

\subsection{Full solution}
\label{single_layer_full_solution}

Here we go beyond the quasi-static approximation and solve the radiated electric field.
According to Eq.~(\ref{Maxwell_electric_field}), the oscillating electromagnetic field is the source for radiating microwaves in space.  Taking  the curl of the first equation in Eq.~(\ref{Maxwell_electric_field}), the electric field of frequency $\omega$ obeys
\begin{align}
    \nabla^2{\bf E}+\varepsilon_0\mu_0\omega^2{\bf E}=-i\omega\mu_0\nabla\times {\bf M}.
    \label{single_electric_field}
\end{align}
Such a radiation process is governed by the oscillating ``magnetization current" ${\bf J}_M=\nabla\times {\bf M}$,  which is analogous to the radiation caused by the normal oscillating charge current~\cite{Jackson}.

Via the Green function technique~\cite{Jackson}, Eq.~(\ref{single_electric_field}) has the solution 
\begin{align}
    {\bf E}({\bf r})=\dfrac{i\mu_0 \omega}{4\pi}\int \dfrac{[\nabla'\times {\bf M}({\bf r'})]e^{i k|{\bf r-r'}|}}{|{\bf r-r'}|}d{\bf r'},
\end{align}
where $k=\omega/c$ is the wave number of microwaves. 
Since only the $x$ and $y$ components of ${\bf M}$ oscillate with frequency $\omega$ and  $\bf M$ is uniform inside the ferromagnetic layer,  $(\nabla\times {\bf M})_{x,y}=0$ in all space, leading to $E_x=E_y=0$ and
\begin{align}
    E_z(x)=\dfrac{i\mu_0 \omega}{4\pi}\int \dfrac{[\partial_{x'}M_y({\bf r'})]e^{i k|{\bf r-r'}|}}{|{\bf r-r'}|}d{\bf r'}.
\end{align}
Using Weyl identity~\cite{Yu_chirality}
\begin{align}
\frac{e^{ik|{\bf r}-{\bf r'}|}}{|{\bf r}-{\bf r}'|}=\int dk_z'dk_y'\frac{ie^{i k_z'(z-z')+i k_y'(y-y')}e^{i\sqrt{k^2-k_z'^2-k_y'^2} |x-x'|}}{2 \pi\sqrt{k^2-k_z'^2-k_y'^2}},
\label{Weyl_identity}
\end{align}
we obtain the electric field 
\begin{align}
      E_z=\dfrac{\mu_0 \omega M_y}{2 k}\begin{cases}
          e^{-ik(x-d_F)}-e^{ik(x+d_F)},  & -d_F<x<d_F \\
      e^{ik(x-d_F)}-e^{ik(x+d_F)},  &     x>d_F\\
       e^{-ik(x-d_F)}-e^{-ik(x+d_F)},  &     x<-d_F
    \end{cases}.
\label{full_solution_single_layer}
\end{align}
From Eq.~(\ref{Maxwell_electric_field}), we find the magnetic induction $B_x=0$, $B_z=\mu_0(H_0+M_0)$ is static, and $B_y=-\partial_xE_z/(i\omega)$ follows
\begin{align}
      B_y=\dfrac{\mu_0  M_y}{2}\begin{cases}
    e^{ik(x+d_F)}+e^{-ik(x-d_F)},  & -d_F<x<d_F \\
      e^{ik(x+d_F)}-e^{ik(x-d_F)},  &     x>d_F\\
       -e^{-ik(x+d_F)}+e^{-ik(x-d_F)},  &     x<-d_F
    \end{cases}.
    \label{magnetic_field}
\end{align}

\begin{figure}[h]
    \centering
    \includegraphics[width=0.9\linewidth]{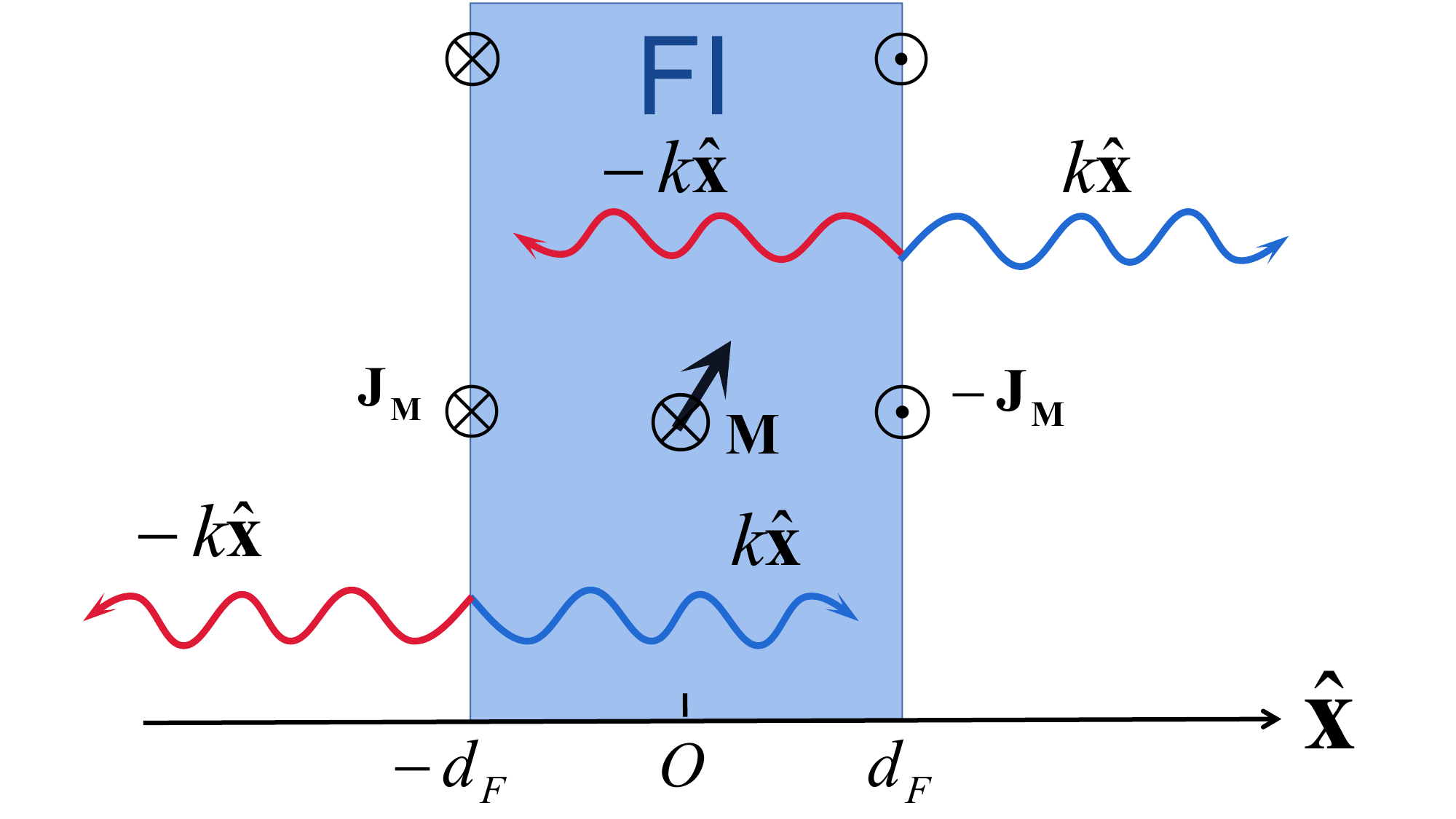}
   \caption{Electric field radiated from the surface magnetization current at the two surfaces of the magnetic insulator.}
    \label{Radiation}
\end{figure}

We can understand the radiated electric field (\ref{full_solution_single_layer}) well via the oscillating ``magnetization current" ${\bf J}_M$. For the uniform magnetization precession, ${\bf J}_M$ is located at the surfaces of the ferromagnetic insulator, i.e., the dynamic component
\begin{align}
{\bf J}_{M}(x)=[\delta(x+d_F)-\delta(x-d_F)]M_y\hat{\bf z}\propto M_y
\label{magnetization_current}
\end{align}
 has the same magnitude but opposite sign at two surfaces $x=\pm d_F$, as illustrated in Fig.~\ref{Radiation}. 
Such oscillating magnetization current then radiates the electromagnetic waves of wave vector $k\hat{\bf x}$ and $-k\hat{\bf x}$ with $k=\omega/c$ into two opposite directions. Due to the opposite sign of ${\bf J}_M$ at $x=\pm d_F$, the amplitudes of the electric fields radiated by the left and right surfaces are of opposite sign $E_L=-E_R\equiv E_0\propto M_y$. 
At the right-hand side of the sample, i.e., $x>d_F$, the propagation phases of the radiated electric field from the left and right surfaces are $k(x+d_F)$ and $k(x-d_F)$, respectively, resulting in a net electric field $E=E_0(e^{ik(x+d_F)}-e^{ik(x-d_F)})$. Similarly, when $x<-d_F$, $E=E_0(e^{-ik(x-d_F)}-e^{-ik(x+d_F)})$. These recover exactly the solution (\ref{full_solution_single_layer}).

With the full solutions (\ref{full_solution_single_layer}) and (\ref{magnetic_field}), we are allowed to calculate the radiation damping of the FMR due to the energy radiated out of the magnetic insulator.  According to Eq.~(\ref{magnetic_field}), the radiated magnetic field inside the magnetic insulating film
\begin{align}
H^r_x&=-M_x,\nonumber\\
H^r_y&=i\frac{\omega d_F M_y}{c}=-\frac{d_F}{c}\frac{dM_y}{dt}
\end{align}
drives the magnetization, leading to the linearized LLG equation 
\begin{align}
  -i\omega M_x+\mu_0 \gamma M_y H_0&=i (\alpha_G+\alpha_R) \omega M_y,\nonumber\\
	i\omega M_y+\mu_0 \gamma H_0 M_x &=-\mu_0 \gamma M_0 M_x+i\alpha_G \omega M_x,
 \label{LLG_single_layer}
\end{align}
where the damping coefficient contributed by the radiation reads
\begin{align}
    \alpha_R=\mu_0\gamma M_0d_F/c.
\end{align}
It is negligibly small: for the YIG film of thickness  $2d_F=120$~nm and $\mu_0M_0=0.2$~T~\cite{magnon_conductivity,YIG_parameter}, $\alpha_R\approx7.3\times 10^{-6}\ll\alpha_G\sim 5\times10^{-4}$. 
However, the radiation damping is enhanced with thicker films.

We are interested in the field near the ferromagnet with a distance $\sim \lambda$.
In ferromagnetic insulators, $\omega\sim2\pi \times 4$~GHz~\cite{superconductor_gating_theory}, and $\lambda\sim100$~nm for conventional superconductors, so $k\lambda\sim 10^{-5}
\ll 1$. 
When $kx\rightarrow0$, we have
\begin{align}
      E_z(x)=\begin{cases}
          -i\mu_0 \omega M_yx,  & -d_F<x<d_F \\
      -i\mu_0 \omega M_yd_F,  &     x>d_F\\
       i\mu_0 \omega M_yd_F,  &     x<-d_F
    \end{cases}, 
    \label{electric_field_quasistatic}
\end{align}
as plotted in Fig.~\ref{electric_field}(a) for a snapshot. The magnetic induction 
\begin{align}
      B_y(x)=\begin{cases}
         \mu_0  M_y,  & -d_F<x<d_F \\
      0,  &     x>d_F\\
      0,  &     x<-d_F
    \end{cases}
\end{align}
recovers to the results from quasi-static approximation~\cite{Rezende} with vanishing magnetic field $H_y$ outside of the ferromagnet.

\subsection{Quasi-static approximation}

The above analysis implies that when focusing on the near-field limit, we may apply the quasi-static approximation that sets $\nabla\times {\bf H}=0$ in Eq.~(\ref{Maxwell_electric_field}). 
 When focusing on the FMR case, $\bf E$ is translation invariant in the $y$-$z$ plane. \textit{i.e.,} $\partial_z E_x =0$. Taking the $y$-component of Eq.~(\ref{electric_field}), the oscillation of $B_y$ only generate $E_z$ parallel to the magnetization:
\begin{align}
-\partial_xE_z=i\omega \mu_0 M_y.
\label{electric_field_in_f}
\end{align}
 Integrating along $x$ across the ferromagnet yields
\begin{align}
    E_z(x)=-i\omega \mu_0 M_y(x+d_F)+E_z(x=-d_F).
\end{align}
 Thereby, $E_z$ depends linearly on $x$ inside the ferromagnet. Outside the ferromagnet, 
 \begin{align}
    E_z(x)=-2i\omega \mu_0 M_yd_F+E_z(x=-d_F)
\end{align}
is uniform, which is consistent with the vanished magnetic field $H_{y|\text{outside}}=0$ in the quasi-static approximation.
According to the symmetry, $E_z(x=0)=0$, 
so the electric field
is exactly the same as Eq.~(\ref{electric_field_quasistatic}).

\section{S$|$FI heterostructure}
\label{bilayer}

\begin{figure}[htp]
\centering 
\hspace{2.5cm}\includegraphics[width=0.92\linewidth]{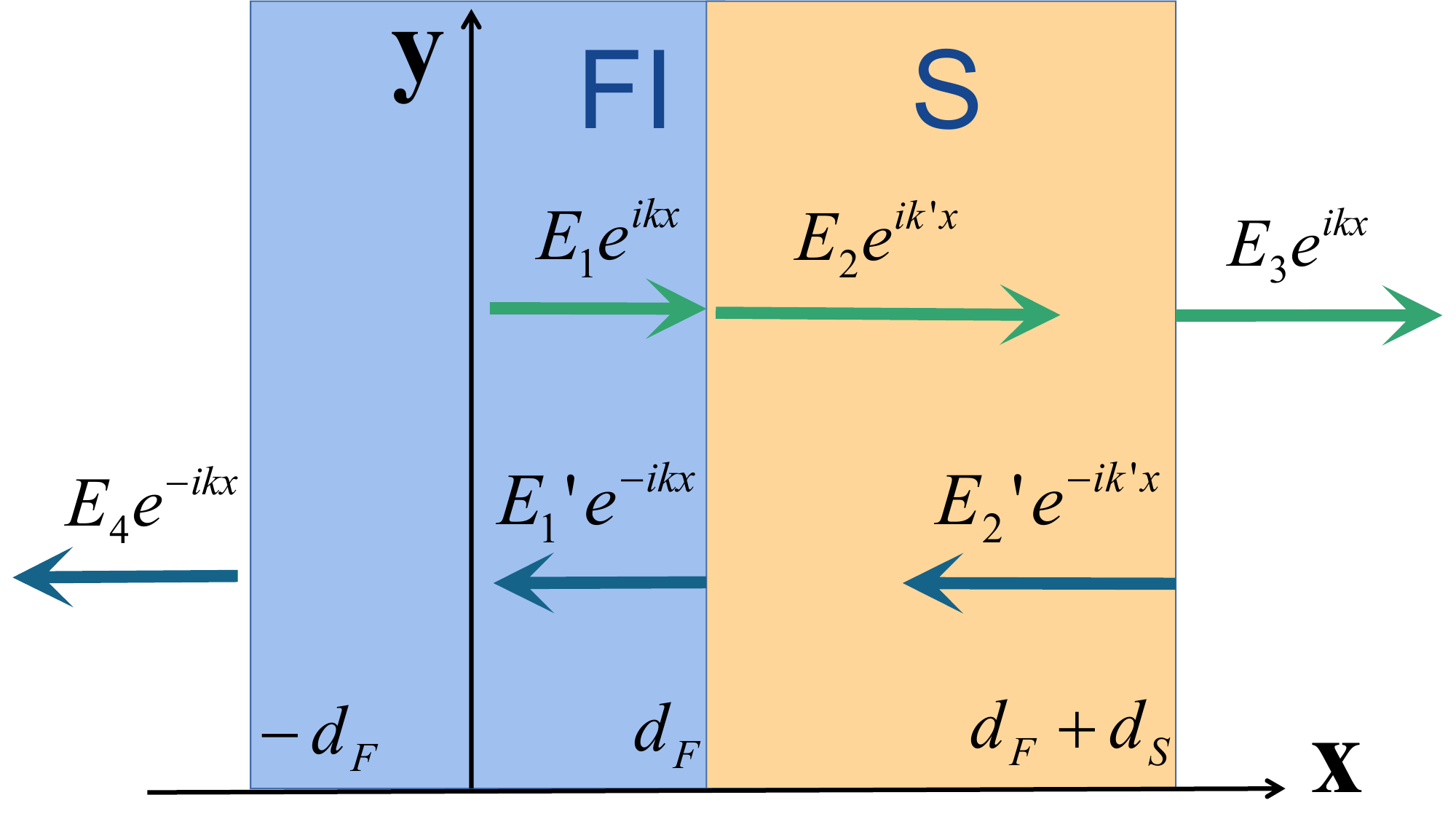}
    \caption{Radiated electric field of the FI$|$S heterostructure.}
    \label{Total_reflection}
\end{figure}

We consider the S$|$FI heterostructure composed of a ferromagnetic film of thickness $2d_F$ and a superconductor of thickness $d_S$, as shown in Fig.~\ref{Total_reflection}. We demonstrate the adjacent superconductors modulate strongly the radiated electric field which explains the absence of the FMR shift in this configuration~\cite{superconductor_gating_exp,CPL_exp}.

\subsection{Full solution}

Inside the ferromagnet, since $\nabla\times{\bf M}=0$ for uniform ${\bf M}$, Eq.~(\ref{single_electric_field}) has the solution $
E_z(x)=E_1 e^{ik x}+E_1'e^{-i k x}$. Inside the superconductor, according to Eqs.~(\ref{electric_field}) and (\ref{London}), the electric field  obeys 
\begin{align}
    \partial_x^2E_z+(\varepsilon_0\mu_0\omega^2-1/\lambda^2)E_z=0,
\end{align}
which has the solution
$E_z(x)=E_2 e^{ik^\prime x}+E_2'e^{-i k^\prime x}$,
where $k^\prime=\sqrt{(\omega/c)^2-1/\lambda^2}\approx i/\lambda$ is purely imaginary with microwave frequencies. For example, with frequency $\omega\sim 2\pi\times 4$~GHz, $k=\omega/c \sim 83.8$~${\rm m}^{-1}$ is much smaller than $1/\lambda \sim 10^{7}~{\rm m}^{-1}$ with London's penetration depth $\lambda\sim 100$~nm. Therefore, due to the Meissner effect, the low-frequency electromagnetic waves no longer propagate but decay in the superconductor. Out of the heterostructure, the electric fields $E_3e^{ikx}$ and $E_4e^{-ikx}$ are radiated. These radiated electric fields are illustrated in Fig.~\ref{Total_reflection}.

The amplitudes $\{E_1,E_1',E_2,E_2',E_3,E_4\}$ are governed by the boundary conditions, i.e., $E_z$ and $H_y$ are continuous at interfaces. The continuous $E_z$ at interface requests 
\begin{align}
    &E_1e^{ik d_F}+E_1^\prime e^{-i k d_F}= E_2e^{ik^\prime d_F}+E_2^\prime e^{-i k^\prime d_F},\nonumber\\
    &E_2e^{ik^\prime (d_F+d_S)}+E_2^\prime e^{-i k^\prime (d_F+d_S)}=E_3e^{i k (d_F+d_S)},\nonumber\\
    &E_1e^{-ik d_F}+E_1^\prime e^{i k d_F}=E_4e^{i kd_F}.
    \label{electric_boundary}
\end{align}
In the superconductors, $H_y=-1/(i\omega\mu_0)\partial_xE_z$, while in the ferromagnet, $H_y=-1/(i\omega\mu_0)\partial_xE_z-M_y$, so the continuous $H_y$ at interfaces leads to 
\begin{align}
    &k^\prime(E_2e^{ik^\prime d_F}-E_2^\prime e^{-i k^\prime d_F})={k}(E_1e^{ik d_F}-E_1^\prime e^{-i k d_F})\nonumber\\
    &+\omega\mu_0 M_y,\nonumber\\
     &{k^\prime}(E_2e^{ik^\prime (d_F+d_S)}-E_2^\prime e^{-i k^\prime (d_F+d_S)})={k}E_3e^{i k (d_F+d_S)},\nonumber\\
    & k(E_1e^{-ik d_F}-E_1^\prime e^{i k d_F})+\omega\mu_0M_y=-kE_4e^{i kd_F}.
    \label{magnetic_boundary}
\end{align}

Combining Eqs.~(\ref{electric_boundary}) and (\ref{magnetic_boundary}), we obtain all the amplitudes. In the ferromagnetic insulator,  
\begin{align}
    E_z(-d_F<x<d_F)={\cal R}E_0e^{-ik(x-d_F)}
    +E_{\text{single}}(x),
    \label{E_bilayer_in_f}
\end{align}
where the amplitude 
$E_0=-[{\omega\mu_0M_y}/({2k})]\left(e^{2ikd_F}-1\right)$, $E_{\text{single}}(x)$ is the radiated electric field from a single magnetic insulator [Eq.~(\ref{full_solution_single_layer})], and 
\begin{align}
    {\cal R}=\frac{e^{ik^\prime d_S}(k^2-k^{\prime2})+e^{-ik^\prime d_S}(k^{\prime2}-k^2)}{e^{ik^\prime d_S}(k-k^{\prime})^2-e^{-ik^\prime d_S}(k+k^{\prime})^2}
\end{align}
is the reflection coefficient of the electric field at the superconductor surface.

\begin{figure}[h]
    \centering
    \includegraphics[width=0.86\linewidth]{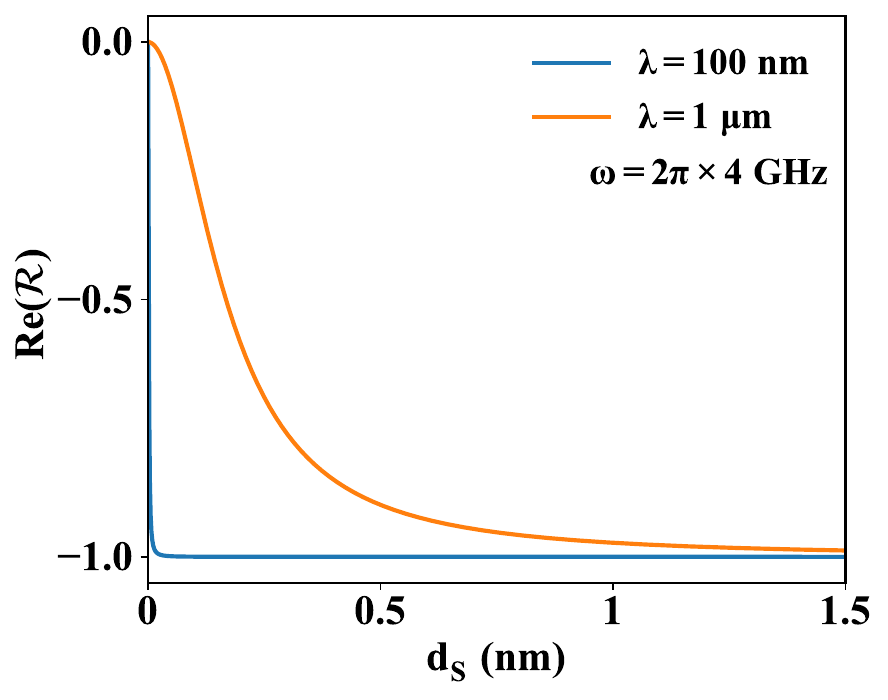}
    \caption{Reflection coefficient Re($\cal R$) as a function of the superconductor thickness $d_S$ with different London's penetration depth $\lambda=100$~nm and $1$~$\mu$m. We take the frequency $\omega=2\pi\times 4$~GHz.}
    \label{reflection_coefficient}
\end{figure}

We plot the dependence of $\cal R$ on the superconductor thickness $d_S$ in Fig.~\ref{reflection_coefficient} with different London's penetration depth $\lambda$ under the frequency $\omega\sim 2\pi \times 4$~GHz. The reflection coefficient saturates to ${\cal R}\rightarrow-1$ when $d_S>0.1$~nm, but is reduced to 0 when $d_S\rightarrow0$, recovering the solution (\ref{full_solution_single_layer}) of the single layer case.
We conclude that even with a small $d_S\ll \lambda$, since $|k|=\omega/c$ is much smaller than $|k'|\approx 1/\lambda$ when $\omega\sim 2\pi \times 4$~GHz, ${\cal R}\rightarrow -1$. This implies the total reflection of the electric fields at the FI$|$S interface even with an ultrathin conventional superconductor layer. As shown below, this indicates the absence of FMR shift in all the available experiments with thick superconductors~\cite{superconductor_gating_exp,CPL_exp}.

Inside the superconductor, 
\begin{align}
    &E_z(d_F<x<d_F+d_S)=\frac{2kE_0}{e^{ik^\prime d_S}(k-k^{\prime})^2-e^{-ik^\prime d_S}(k+k^{\prime})^2}\nonumber\\
    &\times\left((k-k^\prime)e^{-ik^\prime(x-d_F+d_S)}-(k+k^\prime)e^{ik^\prime(x-d_F-d_S)}\right),
\end{align}
which is indeed very weak since $|k|\ll |k'|$.
Out of the heterostructure, 
\begin{align}
  E_z=\begin{cases}
   \dfrac{-4kk^\prime E_0e^{ik(x-d_F-d_S)}}{e^{ik^\prime d_S}(k-k^{\prime})^2-e^{-ik^\prime d_S}(k+k^{\prime})^2},x>d_F+d_S\\
   ~\\
   {\cal R}E_0e^{-ik(x-d_F)}+E_{\text{single}}(x),~x<-d_F
  \end{cases}. 
\end{align} 
At low frequencies and near the heterostructure, $kx\rightarrow0$, $kd_F\rightarrow 0$, and $kd_S\rightarrow 0$, so the electric fields
\begin{align}
    E_z(x)=\begin{cases}
        0,& x>d_F\\
        -i\omega\mu_0M_y(x-d_F), &-d_F<x<d_F\\
        2i\omega\mu_0M_yd_F, &x<-d_F
    \end{cases},
    \label{full_solution_bilayer}
\end{align}
which is illustrated  in Fig.~\ref{electric_field}(b) for a snapshot.
The electric field vanishes in the superconductor due to the total reflection with a $\pi$-phase shift ${\cal R}=-1$ that generates no supercurrent and thereby leads to no modulation on the FMR.

\subsection{Quasi-static approximation}
\label{quasi_static_bilayer}

The full solution clearly shows the 
absence of electric fields at the superconductor side of the S$|$FI heterostructure, which can be well understood within the quasi-static approximation $\nabla\times{\bf H}=0$ or ${\bf J}_s$.  
Assuming $E_z(x=d_F)=\tilde{E}_0$ at the FI$|$S interface, according to Eq.~(\ref{electric_field_in_s}) the electric field in the adjacent superconductor
\begin{align}
    E_z(x)=\tilde{E}_0\dfrac{\cosh{((x-d_S-d_F)/\lambda)}}{\cosh{(d_S/\lambda)}}
\end{align}
drives the supercurrent. For a thin superconducting film of thickness $O(\lambda)$, we are allowed to take an average of the supercurrent $J_{s,z}=[J_{s,z}(x=d_F)+J_{s,z}(x=d_F+d_S)]/2$, and from the first equation of Eq.~(\ref{London})
\begin{align}
    J_{s,z}=\dfrac{i}{\mu_0\omega\lambda^2}\tilde{E}_0\dfrac{1+\cosh(d_S/\lambda)}{2\cosh(d_S/\lambda)}.
\end{align}
The supercurrents generate the vector potential (\ref{vector_potential})
and the Oersted magnetic field according to ${H}_y=-\partial_xA_z/\mu_0$. Taking $k=0$ at low frequencies in the Weyl identity (\ref{Weyl_identity}), i.e.,~\cite{Yu_chirality} 
\begin{align} 	\dfrac{1}{|{\bf r}-{\bf r}'|}=\int dk_x'dk_y'\dfrac{e^{i k_x'(x-x')+i k_y'(y-y')}e^{-\sqrt{k_x'^2+k_y'^2} |z-z'|}}{2 \pi\sqrt{k_x'^2+k_y'^2}},
\label{Weyl_identity_2}
\end{align}
we obtain the Oersted magnetic field generated by the supercurrents
\begin{align}
	H_{s,y}(x)=\left\{
	\begin{array}{cc}
	d_S{{ J}}_{s,z}/2,	& x>d_F+d_S \\
	-d_S{{ J}}_{s,z}/2,	&~ x<d_F 
	\end{array}
	\right. .\label{hy-F|S}
\end{align}
However, constant $H_{s,y}$ independent of $x$ should vanish \textit{out of the heterostructure}  within the quasi-static approximation since a constant magnetic field renders the radiated electric field divergent, which requests $J_{s,z}=0$ when $d_S\ne 0$ and $E_z(x>d_F)=0$. Since the electric field is continuous at interfaces, $E_z(x=d_F)=\tilde{E}_0=0$ and according to Eq.~(\ref{electric_field_in_f}) $E_z(x=-d_F)=2id_F\omega \mu_0M_y$. These simple calculations thereby capture precisely the key physics of the full solution (\ref{full_solution_bilayer}).

\section{S$|$FI$|$S heterostructure}
\label{trilayer}

Further, we consider the S$|$FI$|$S heterostructure as illustrated in  Fig.~\ref{model2} composed of the ferromagnetic insulator of thickness $2d_F$ and two adjacent superconductor films of thickness $d_S$ and $d_S'$, respectively.
In comparison to that of the S$|$FI bilayer, the distribution of the electric field in S$|$FI$|$S heterostructure changes much due to its back-and-forth reflection by the superconductors, as addressed in this section.

\subsection{Full solution}

\begin{figure}[htp]
\centering 
\hspace{2.5cm}\includegraphics[width=0.92\linewidth]{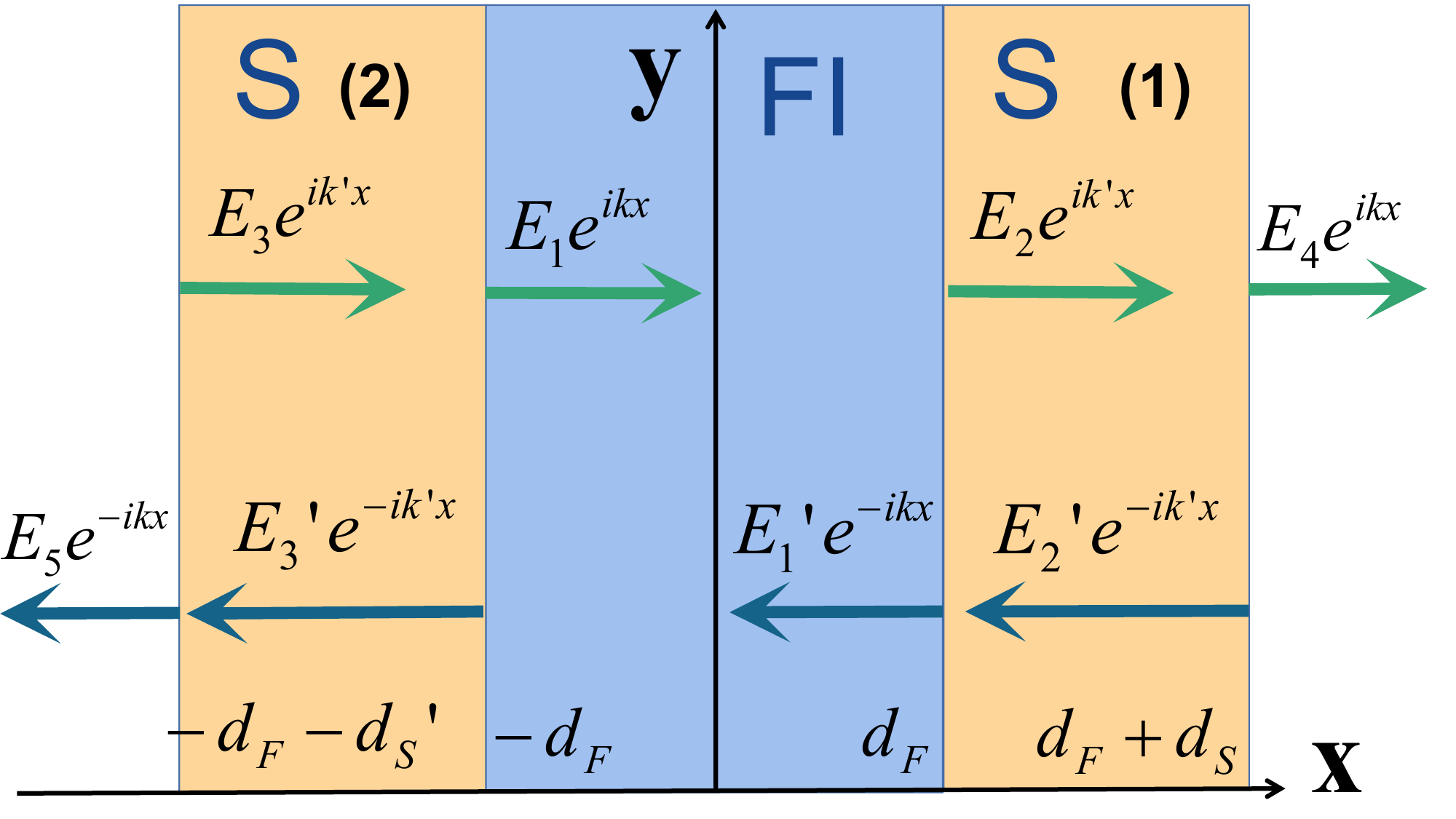}
    \caption{Radiated electric field of the S$|$FI$|$S heterostructure.}
    \label{SFS_reflection}
\end{figure}

Similar to the S$|$FI  heterostructure, inside the ferromagnet, $E_z(x)=E_1 e^{ikx}+E_1^\prime e^{-ikx}$; in the superconductor ``1'', $E_z(x)=E_2e^{ik'x}+E_2'e^{-ik'x}$; and in the superconductor ``2'', $E_z(x)=E_3e^{ik'x}+E_3'e^{-ik'x}$. Out of the heterostructure, the electric fields $E_4e^{ikx}$ and $E_5e^{-ikx}$ are radiated. These electric fields are illustrated in Fig.~\ref{SFS_reflection}.

The amplitudes $\{E_1,E_1',E_2,E_2',E_3,E_3',E_4,E_5\}$ are governed by the boundary conditions. The continuous $E_z$ at interfaces requests
\begin{align}
    &E_1e^{ikd_F}+E_1'e^{-ikd_F}=E_2e^{ik'd_F}+E_2'e^{-ik'd_F},\nonumber\\
    &E_1e^{-ikd_F}+E_1'e^{ikd_F}=E_3e^{-ik'd_F}+E_3'e^{ik'd_F},\nonumber\\
    &E_2e^{ik'(d_F+d_S)}+E_2'e^{-ik'(d_F+d_S)}=E_4e^{ik(d_F+d_S)},\nonumber\\
    &E_3e^{-ik'(d_F+d_S')}+E_3'e^{ik'(d_F+d_S')}=E_5e^{ik(d_F+d_S')},
    \label{electric_boundary_SFS}
\end{align}
and the continuous $H_y$ at interfaces leads to 
\begin{align}
    &k'(E_2e^{ik'd_F}-E_2'e^{-ik'd_F})=k(E_1e^{ikd_F}-E_1'e^{-ikd_F})\nonumber\\
    &+\omega\mu_0M_y,\nonumber\\
    &k'(E_3e^{-ik'd_F}-E_3'e^{ik'd_F})=k(E_1e^{-ikd_F}-E_1'e^{ikd_F})\nonumber\\
    &+\omega\mu_0M_y,\nonumber\\
    &k'(E_2e^{ik'(d_F+d_S)}-E_2'e^{-ik'(d_F+d_S)})=kE_4e^{ik(d_F+d_S)},\nonumber\\
   & k'(E_3e^{-ik'(d_F+d_S')}-E_3'e^{ik'(d_F+d_S')})=-kE_5e^{ik(d_F+d_S')}.
   \label{magnetic_boundary_SFS}
\end{align}

Combining Eqs.~(\ref{electric_boundary_SFS}) and (\ref{magnetic_boundary_SFS}), we obtain the electric-field distribution. In particular, when $d_S=d_S'$, in the ferromagnetic film,
\begin{align}
    E_z(|x|<d_F)=\frac{-\omega\mu_0M_y\sinh{(ikx)}}{k\cosh{(ikd_F)}-k'f(u)\sinh{(ikd_F)}},
    \label{symmetric_E}
\end{align}
where $u=-[(k+k')/(k-k')]\exp(-2ik'd_S)$ and 
\begin{align}
    f(u)=\frac{u-1}{u+1}=\frac{k'\sinh{(ik'd_S)}-k\cosh{(ik'd_S)}}{k\sinh{(ik'd_S)}-k'\cosh{(ik'd_S)}}.
\end{align}
In the superconductor ``1", 
\begin{align}
    &E_z(d_F<x<d_F+d_S)\nonumber\\
    &=\frac{-\omega\mu_0M_y(ue^{ik'(x-d_F)}+e^{-ik'(x-d_F)})}{k(1+u)\coth(ikd_F)-k'(u-1)},
    \label{Electric_field_in_(1)}
\end{align}
and in the superconductor ``2",
\begin{align}
    &E_z(-d_F-d_S<x<-d_F)\nonumber\\
    &=\frac{\omega\mu_0M_y(ue^{-ik'(x+d_F)}+e^{ik'(x+d_F)})}{k(1+u)\coth(ikd_F)-k'(u-1)}.
\end{align}
They both exist, and $E_z(x=-d_F)$ and $E_z(x=d_F)$ are opposite. This feature may be understood from the magnetic dipole radiation: since the magnetization current ${\bf J}_M$ (\ref{magnetization_current}) is opposite at the two surfaces $x=\pm d_F$ of the magnetic film, the amplitudes of the electric fields radiated by the two surfaces $x=\pm d_F$ are of opposite sign, which launches to superconductors and drives the opposite supercurrents in them.

Out of the heterostructure,
\begin{align}
    &E_z(x>d_F+d_S)\nonumber\\
    &=\frac{-\omega\mu_0M_y(ue^{ik'd_S}+e^{-ik'd_S})}{k(1+u)\coth(ikd_F)-k'(u-1)}e^{ikx},\nonumber\\
    &E_z(x<-d_F-d_S)\nonumber\\
    &=\frac{\omega\mu_0M_y(ue^{ik'd_S}+e^{-ik'd_S})}{k(1+u)\coth(ikd_F)-k'(u-1)}e^{-ikx},
\end{align}
which, when far away from the heterostructure, is reduced to a simpler form 
\begin{align}
    E_z(x)\approx&\frac{i\omega\mu_0d_F\lambda M_y}{\lambda\cosh{(d_S/\lambda)}+d_F\sinh{(d_S/\lambda)}}\nonumber\\
    \times&\begin{cases}
 -e^{ikx}, & x\gg d_F+d_S\\
e^{-ikx},& x\ll -(d_F+d_S)
    \end{cases}.
\end{align}
The radiation out of the heterostructure is then completely suppressed when $d_S\gg \lambda$.

\begin{figure}[htp]
    \centering
    \includegraphics[width=0.81\linewidth]{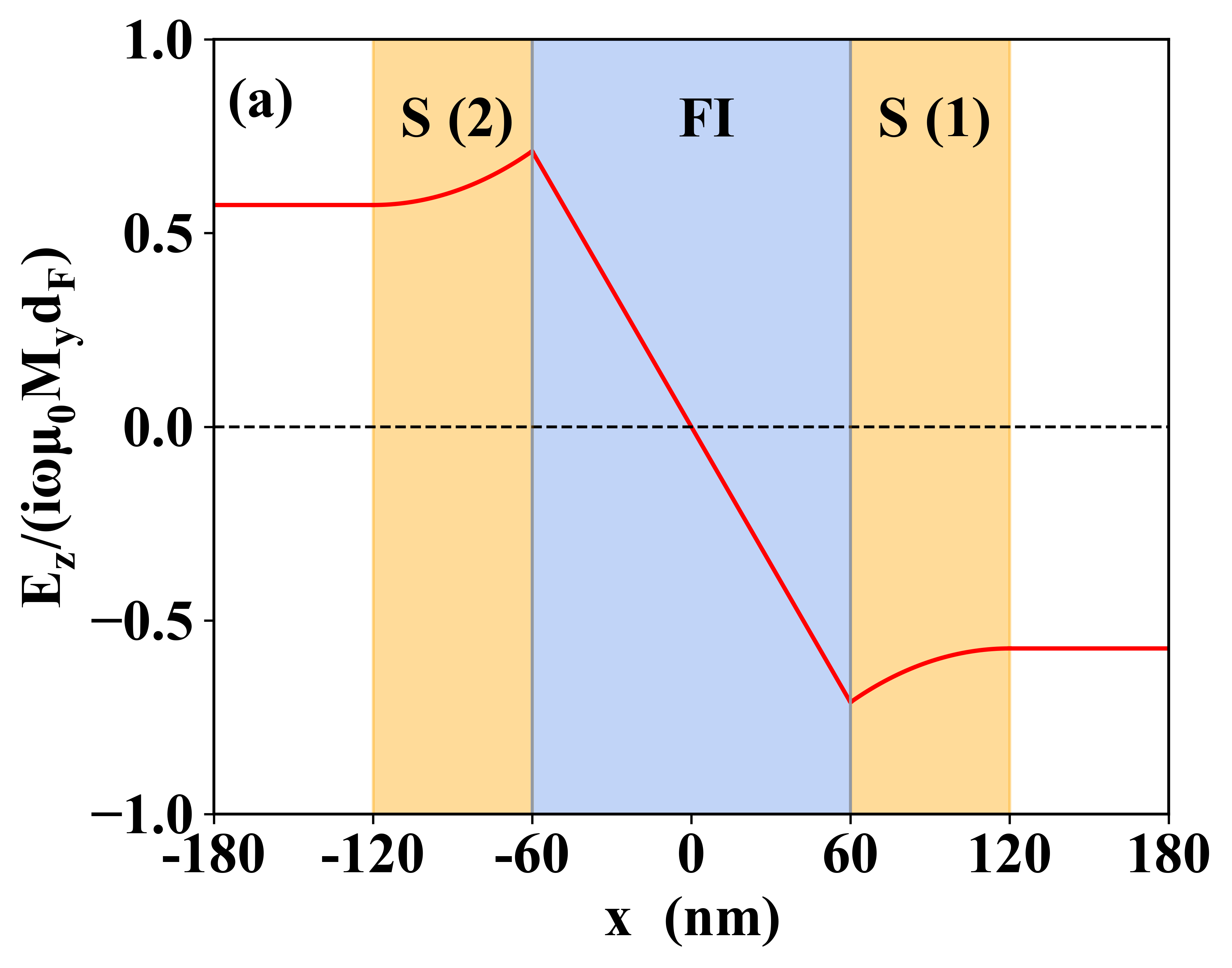}
    \includegraphics[width=0.81\linewidth]{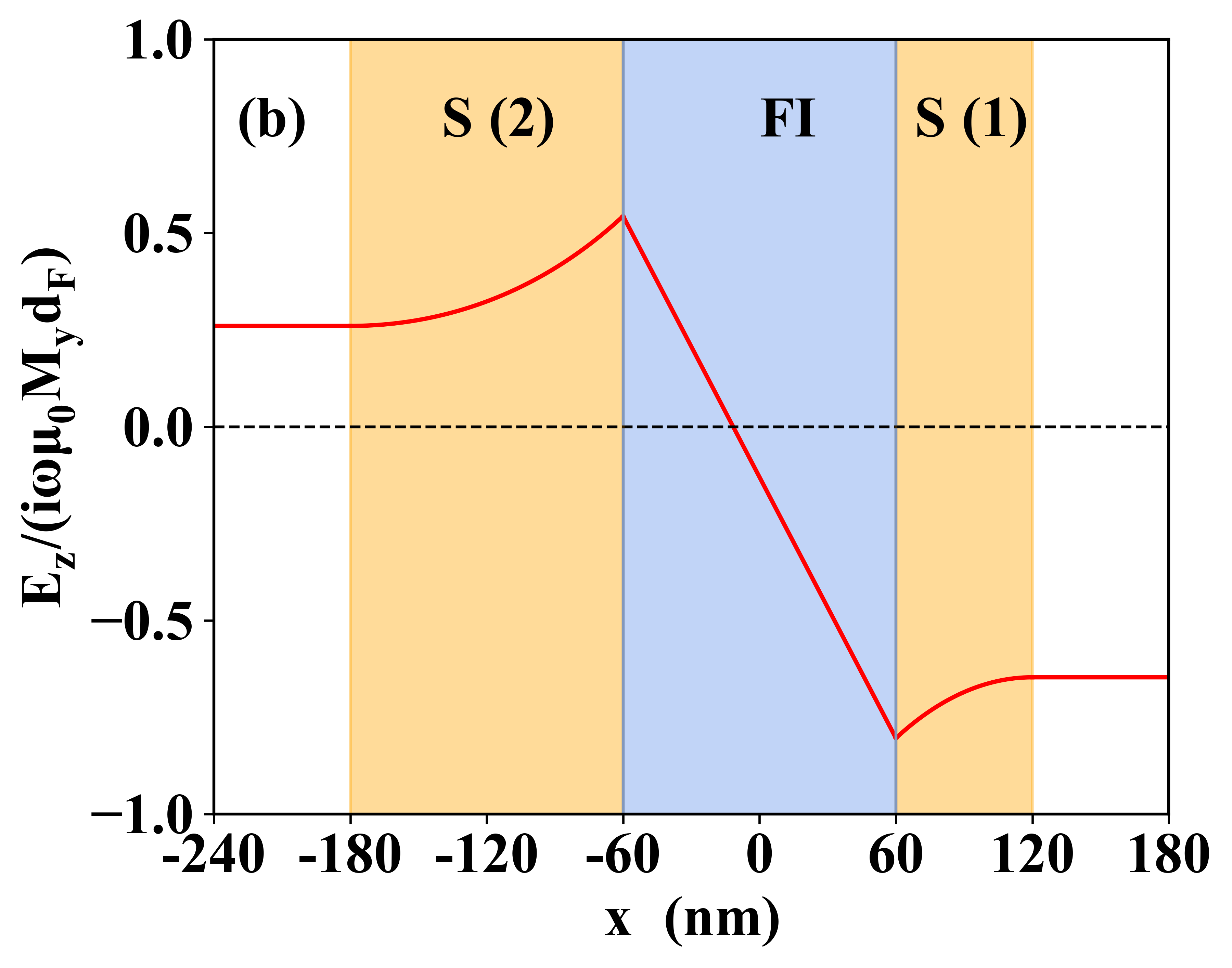}
    \caption{Distribution of electric fields  in symmetric $d_S=d_S'=60$~{nm} [(a)] and asymmetric $d_S'=2d_S=120$~nm [(b)] S$|$FI$|$S heterostructure. The thickness of the ferromagnetic film 2$d_F=120$~nm and London's penetration depth $\lambda(T=0.5T_c)=87.8$~nm.}
    \label{electric_distribution}
\end{figure}

We refer to Appendix~\ref{appendix} for the solution of asymmetric configuration.
We illustrate in Fig.~\ref{electric_distribution} the distribution of the electric fields ${\rm Re}(E_z/(i \omega\mu_0M_yd_F))$ at $T=0.5T_c=5.5$~K in the symmetric $d_S'=d_S=60$~nm and asymmetric $d_S'=2d_S=120$~nm S$|$FI$|$S heterostructure, respectively, in the near-field limit. For NbN, $T_c=11$~K, the London penetration depth $\lambda(T=0)=85$~nm~\cite{NbN_lambda} and $\lambda(T=0.5T_c)=87.8$~nm. The fields are opposite at the two superconductors in the symmetric heterostructure but are skewed when $d_S\ne d_S'$. These fields carrying energy are radiated out in the far zone~\cite{Jackson}. When the superconductors are sufficiently thick $\{d_S,d_S'\}\gg \lambda$, these electric fields are confined between them, which corresponds to an excellent waveguide with small size~\cite{Swihart}.

\subsection{Ultrastrong interaction between Kittel magnon and Cooper-pair supercurrent}

Above we address that the dynamics of magnetization ${\bf M}$  generates $H^r_{y}$ via the backaction of superconductors, which, in turn, drives ${\bf M}$ in the ferromagnet, imposing a self-consistent problem that is solved by combining the Landau-Lifshitz and Maxwell's equations. 
In other words, the precession of the magnetization radiates the electric field that drives the supercurrent in the superconductor via microscopically generating the center-of-mass momentum of the Cooper pairs. Such a collective motion of Cooper pairs, i.e., the supercurrent in turn generates the Oersted magnetic field that affects the dynamics of the magnetization, i.e., a shift in its FMR frequency.

Using Eq.~(\ref{symmetric_E}) and  $B_y=-\partial_xE_z/(i\omega)$, we find the radiated magnetic field inside the ferromagnetic insulator of the symmetric S$|$FI$|$S heterostructure
\begin{align}
    H_{y}^r(|x|<d_F)&=\frac{M_yk \cosh{(ikx)}}{k\cosh{(ikd_F)}-k'f(u)\sinh{(ikd_F)}}-M_y,
    \label{radiated_magnetic_field}
\end{align}
which drives the precession of the magnetization. In terms of the (linearized) LLG equation (\ref{Full_LLG_Equation}), we arrive at
\begin{align}
     -i\omega M_x+\mu_0 \gamma M_y H_0&=\mu_0 \gamma M_0H_{y}^{r}+i \alpha_G \omega M_y,\nonumber\\
\mu_0 \gamma H_0 M_x+i\omega M_y &=-\mu_0 \gamma M_0 M_{x}+i\alpha_G \omega M_x.
\label{linearized_LLG_Equation}
 \end{align}
We see that the real part of the radiated magnetic field (\ref{radiated_magnetic_field}) is in the same phase of $M_y$, which provides a field-like torque for the magnetization. Retaining the leading order in
$k$, the homogeneous
 \begin{align}
     \Re(H_{y}^r)=-\frac{d_F\tanh(d_S/\lambda)}{\lambda+d_F\tanh(d_S/\lambda)}M_y
 \end{align}
 renormalizes the FMR frequency to be 
 \begin{align}
     \omega_{\rm K}=\mu_0\gamma\sqrt{(H_0+M_0)\left(H_0+\dfrac{d_F\tanh(d_S/\lambda)}{\lambda+d_F\tanh(d_S/\lambda)}M_0\right)},
     \label{FMR_frequency}
 \end{align}
which differs from the bare Kittel frequency $\tilde{\omega}_{\rm K}=\mu_0\gamma\sqrt{H_0(H_0+M_0)}$~\cite{kittel_mode}. When $d_S\gg \lambda$, the solution (\ref{FMR_frequency}) recovers that in Ref.~\cite{Silaev_ultrastrong_coupling}, where an ultrastrong coupling between magnons and microwave photons is predicted in a magnetic insulator when sandwiched by two superconductors of \textit{infinite} thickness.

On the other hand, the imaginary part of the radiated magnetic field is out of phase of $M_y$, which thereby
contributes to a damping-like torque.
Retaining the leading order in $k$, 
\begin{align}
     &\Im(H_y)\approx \frac{M_y k d_F}{\cosh^2\left({d_S}/{\lambda}\right)}
    \left(1+\dfrac{d_F\tanh{(d_S/\lambda)}}{\lambda}\right)^{-2}
     \nonumber
\end{align}
 contributes to a damping coefficient 
 \begin{align}
 \nonumber
     {\alpha_R}&=\dfrac{\mu_0\gamma M_0d_F}{c\cosh^2\left({d_S}/{\lambda}\right)} \left(1+\dfrac{d_F\tanh{(d_S/\lambda)}}{\lambda}\right)^{-2}.
     \nonumber
 \end{align}
 In comparison to a single layer of magnetic insulator (Sec.~\ref{single_layer_full_solution}), the radiation of magnetization is suppressed when shielded by two superconductors, and the radiation damping is expected to be reduced.
 With $d_S=d_F=60$~nm, $\lambda\sim 85$~nm, and $\omega\sim 2\pi \times 4$~GHz, $\alpha_R\approx 2.2\times 10^{-6}$ is indeed smaller than that of a single magnetic insulator ($7.3\times 10^{-6}$). When $d_S\gg \lambda$, $\alpha_R\rightarrow 0$ since no field is radiated out of the S$|$FI$|$S heterostructure.

The general solution of $\omega_{\rm K}$ [Eq.~(\ref{general_omega_K})] and $\alpha_R$ [Eq.~(\ref{general_alpha_R})] in the asymmetric S$|$FI$|$S heterostructure is calculated in Appendix.~\ref{appendix}. In Appendix~\ref{appendix_B}, we calculate them with the quasistatic approximation.

To show the FMR shift, we assume an oscillating magnetic field 
$\tilde{H}e^{-i\omega_0 t}\hat{\bf y}$ of frequency $\omega_0$ applied along the $\bf \hat{y}$-direction (the associated microwave electric field is along the normal $\hat{\bf x}$-direction). 
The wavelength of this microwave is much larger than the thickness of the heterostructure, so it can be treated as uniform across the heterostructure thickness. 
It can penetrate the superconductor easily when $\{d_S,d_S'\}\sim \lambda$. With the wave vector (along $\hat{\bf z}$)
parallel to the film, it only excites $\bf M$ in the ferromagnet but does not drive the superconductor.

Including the external pump field $\tilde{H}e^{-i \omega_0 t}\hat{\bf y}$  into the LLG equation (\ref{linearized_LLG_Equation}),
we find when $\alpha_G\ll1$ 
\begin{align}
    M_y&=\dfrac{\mu_0^2\gamma^2M_0(H_0+M_0)}{\omega_{\rm K}^2-\omega_0^2-i \Gamma}\tilde{H},
	\nonumber\\
	M_x&=-i M_y\left[ \dfrac{\omega_0}{\mu_0\gamma(H_0+M_0)}+\dfrac{i \alpha_G \omega_0^2}{(\mu_0\gamma(H_0+M_0))^2}\right],
\end{align}
where
\begin{align}
    \Gamma=\dfrac{\alpha_G\omega_0^3}{\mu_0\gamma(H_0+M_0)}+\mu_0\gamma(H_0+M_0)(\alpha_G+\alpha_R)\omega_0.
\end{align}
From Eq.~(\ref{Electric_field_in_(1)}), we find the average electric field $E_z=[E_z(x=d_F)+E_z(x=d_F+d_S)]/2$ in the thin superconductor ``1" as
\begin{align}
    E_z^{(1)}=&-\dfrac{\tilde{H}}{2}\dfrac{\omega\mu_0(u+1+ue^{i k' d_S}+e^{-i k' d_S})}{k(1+u)\coth(i k d_F)-k'(u-1)}\nonumber\\
    \times&\dfrac{\mu_0^2\gamma^2M_0(H_0+M_0)}{\omega_{\rm K}^2-\omega_0^2-i \Gamma}.
\end{align}
From Eq.~(\ref{London}), the corresponding average suppercurrent inside the superconductor is
\begin{align}
    J_z^{(1)}=&-\dfrac{i\tilde{H}}{2\lambda^2}\dfrac{u+1+ue^{i k' d_S}+e^{-i k' d_S}}{k(1+u)\coth(i k d_F)-k'(u-1)}\nonumber\\
    \times&
    \dfrac{\mu_0^2\gamma^2M_0(H_0+M_0)}{\omega_{\rm K}^2-\omega_0^2-i \Gamma}.
\end{align}

\begin{figure}[htp]
	\centering	\includegraphics[width=0.82\linewidth]{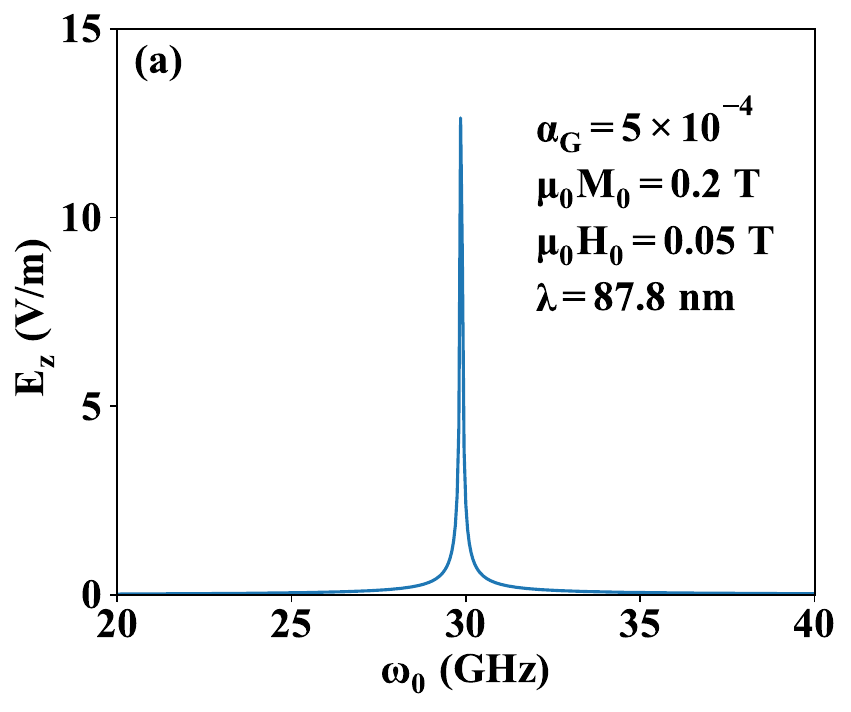}
\includegraphics[width=0.82\linewidth]{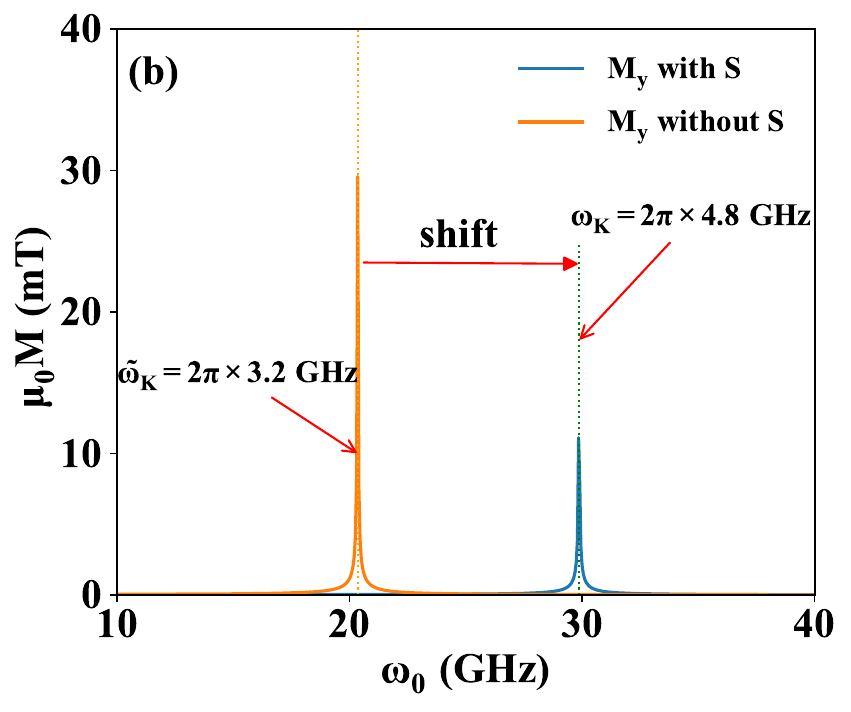}
\includegraphics[width=0.82\linewidth]{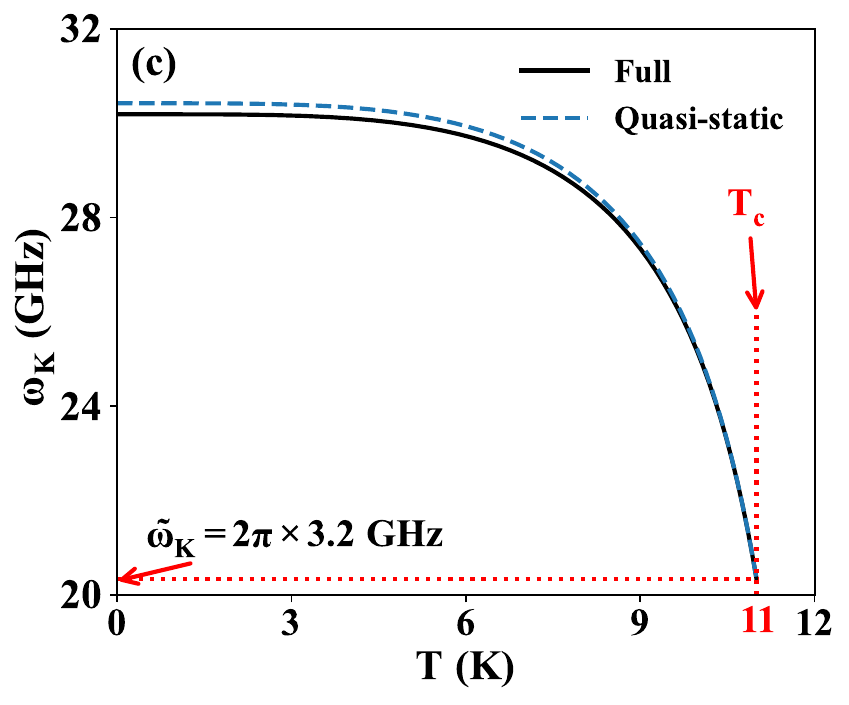}
\caption{FMR spectra with the excitation field $\mu_0\tilde{H}=0.01$~mT. In (a) and (b), we use the temperature $T=0.5T_c=5.5$~K. (a) plots the excited electric field amplitude in (one of) the superconductors in the symmetric S$|$FI$|$S heterostructure. The amplitude of the resonance electric field $E_z\sim 14$~V/m. (b) is the excited amplitudes of the magnetization $M_y$
with and without two adjacent superconductors. $M_x\approx0.6M_y$ and $0.5M_y$ with and without the superconductors. The frequency shift is as large as $2\pi\times 1.6$~GHz$\sim \tilde{\omega}_{\rm K}/2$. (c) is the temperature dependence of FMR frequency $\omega_{\rm K}$ by solutions with the full calculation (black line) and quasi-static approximation (dashed line). The bare FMR frequency $\tilde{\omega}_{\rm K}=2\pi \times 3.2$~GHz.}
	\label{shift_results}
\end{figure}

We illustrate the numerical results considering a yttrium iron garnet (YIG) film of thickness $2d_F=120$~nm sandwiched by two NbN superconductors of thickness $d_S=d_S'=60$~nm.
Insulating EuS thin magnetic film~\cite{EuS_thin_film,thin_film} is also a possible candidate to test our prediction. For YIG, $\mu_0M_0=0.2$~T and $\alpha_G=5\times 10^{-4}$~\cite{magnon_conductivity,YIG_parameter}. We use $\lambda(T=0.5T_c)=87.8$~nm for  NbN~\cite{NbN_lambda} .  
We take the bias field $\mu_0H_0=0.05$~T and the excitation field $\mu_0\tilde{H}=0.01$~mT.
Figure~\ref{shift_results} shows the radiated electric field in (one of) the superconductors and the excited amplitudes of $\bf M$ as a function of the excitation frequency $\omega_0$. The frequency shift is $2\pi\times 1.6$~GHz, comparable to half of the bare FMR frequency $\tilde{\omega}_{\rm K}=2\pi\times 3.2$~GHz, corresponding to the decrease of the resonant magnetic field as large as 55~mT. This demonstrates the potential to achieve ultrastrong interaction between magnons and Cooper-pair supercurrent even with magnetic insulators.

Before we address the temperature dependence of the frequency shift, we first show that the normal current mainly provides additional damping to the FMR with a tiny frequency shift even when $T\rightarrow T_c$.  We estimate the contribution of the normal current via the two-fluid model with the conductivity at low frequencies~\cite{London_equation}
\begin{align}
    \Tilde{\sigma}(\omega)\approx\dfrac{\rho_ne^2\tau}{m_e}+i\dfrac{\rho_se^2}{m_e}\dfrac{1}{\omega}=\sigma_n+i\frac{1}{\omega\mu_0\lambda^2}.
    \label{conductivity}
\end{align}
where $\tau$ is the relaxation time of electrons and $\rho_n$ $(\rho_s)$ is the normal fluid (superfluid) density. $\rho_n$ equals to the electron density $n_e$ when $T>T_c$. 
Incorporating the conductivity (\ref{conductivity}) into Maxwell's equation, the radiated magnetic field contributed by both the normal and supercurrents in the symmetric S$|$FI$|$S heterostructure (to the leading order of $k$) reads
\begin{align}
   \tilde{H}_{y}&=M_y\frac{i \tilde{k}d_F(\tilde{k}\tanh{(i \Tilde{k}d_S)}-k)}{\tanh{(i \Tilde{k}d_S)}(k-i \tilde{k}^2d_F)+\tilde{k}(ikd_F-1)},
\end{align}
where $\tilde{k}^2=i\omega\mu_0\sigma_n-1/\lambda^2$, with which we find the FMR frequency and the additional damping coefficient 
\begin{align}
\omega_{\rm K}&=\mu_0\gamma\sqrt{H_0+M_0}
     \sqrt{H_0-M_0\Re(\tilde{H}_y)/M_y},\nonumber\\
     \tilde{\alpha}&={\mu_0\gamma M_0\Im(\tilde{H}_y)}/({\omega_{\rm K}M_y}).
\end{align}
When $T\rightarrow T_c$, with $\sigma_n\sim 1.1\times 10^6$~$(\Omega\cdot \text{m})^{-1}$ for NbN~\cite{cooper_pair_density}, $d_S=d_F=60$~nm, and $\omega\sim 2\pi\times 4$~GHz, we find the frequency shift $\delta\omega=\omega_{\rm K}-\tilde{\omega}_{\rm K}\sim 10^{-5}$~GHz is negligibly small, while the additional damping is considerably large $\tilde{\alpha}\sim 2\times 10^{-4}$ for YIG.

Since the normal current can be disregarded in the frequency shift, we calculate the temperature dependence of the FMR frequency according to Eq.~(\ref{penetration_depth}), as plotted in Fig.~\ref{shift_results}(c) with the same parameters used in Fig.~\ref{shift_results}(a) and (b). When $T\rightarrow0$, the resonance frequency reaches its maximum, while when $T\rightarrow T_c$, the resonance frequency recovers to the Kittel bare frequency since the superconductivity is depleted. We compare the full solution (black line) and the quasi-static solution (dashed line) and find the quasi-static approximation is excellent in all the temperature regimes when $d_S\lesssim \lambda$.

\section{Conclusion and discussion}
\label{conclusion}

Magnetic insulators are ideal candidates for long-range spin transport~\cite{Lenk,Chumak,Grundler,Demidov,Brataas,Barman,Yu_chirality}, strong coupling between magnons and microwaves~\cite{Q_information_4}, and quantum information processing~\cite{Q_information,Q_information_1,Q_information_5,Q_information_3,Q_information_6}, gating which by superconductors may bring new control dimensions.
In comparison to metallic magnets, the mutual proximity effect may differ between magnetic insulators and superconductors, which may be helpful to distinguish different competitive mechanisms~\cite{Bergeret} in future studies. Our model system differs from the \textit{metallic} ferromagnets since there are no electric currents flowing in the insulators that, if large, may affect the field distribution via radiation. 
  
The formulation of the response in the superconductor by London's equation is phenomenological, which, nevertheless, captures the key physics of the interplay between FMR in the magnetic insulator and supercurrent in the superconductor. Some interesting effects, such as the role of impurity and finite correlation length of Cooper pairs, may be not precisely taken into account in the classical London model, however. Our work can be a starting point for an extension to a fully microscopic model in terms of, e.g., the Usadel equation~\cite{Usadel_Equation}, in the future.
    
In conclusion, we analyze the interaction between the Kittel magnons in insulating magnetic film and Cooper-pair supercurrent in superconductors mediated by the radiated electric fields from the magnetization dynamics. Via highlighting the role of the total reflection of the electric fields at the ferromagnet-superconductor interface that are solved beyond the quasi-static approximation, we provide a comprehensive understanding of the absence of the FMR shift in the FI$|$S heterostructure and predict its existence in the S$|$FI$|$S heterostructure with the Meissner screening. The coupling between magnons and Cooper-pair supercurrent is ultrastrong with the frequency shift achieving tens of percent of the bare FMR frequency, which may bring superior advantage in information processing in on-chip magnonics and quantum magnonics.

\begin{acknowledgments}
We gratefully acknowledge Prof.~Guang Yang and Prof.~Lihui Bai 
for many inspiring discussions.
This work is financially supported by the National Natural Science Foundation of China under Grant No.~12374109, and the startup grant of Huazhong University of Science and Technology (Grants  No.~3004012185 and 3004012198). 
\end{acknowledgments}

\begin{appendix}

\section{General solution of $E_z$ in S$|$FI$|$S heterostructure}
 \label{appendix}

Here we list the general solution of $E_z(x)$ in the S$|$FI$|$S heterostructure when $d_S\ne d_S'$ in Fig.~\ref{SFS_reflection}. Inside the ferromagnet,
\begin{align}
    &E_z(-d_F<x<d_F)\nonumber\\
    &=\frac{-\omega\mu_0M_y(Ge^{ikx}+e^{-ikx})}{k(Ge^{ikd_F}-e^{-ikd_F})-k'f(u)(Ge^{ikd_F}+e^{-ikd_F}) },
    \label{asymmetric_E}
\end{align}
    where 
    \begin{align}
        G=-\frac{-2k\sinh(ikd_F)+k'(f(u)e^{-ikd_F}+f(u')e^{ikd_F})  
        }{-2k\sinh(ikd_F)+k'(f(u)e^{ikd_F}+f(u')e^{-ikd_F})},
    \end{align}
    and $u'=-[(k+k')/(k-k')]\exp(-2ik'd_S')$.
In the superconductor ``1'',
\begin{align}
    E_z(d_F<x<d_F+d_S)=\frac{ue^{ik'(x-d_F)}+e^{-ik'(x-d_F)}}{1+u}\nonumber\\
    \times \frac{-\omega\mu_0M_y(Ge^{ikd_F}+e^{-ikd_F})}{
    k(Ge^{ikd_F}-e^{-ikd_F})-k'f(u)(Ge^{ikd_F}+e^{-ikd_F})
    }.
\end{align}
In the superconductor ``2'',
\begin{align}
    &E_z(-d_F-d_S'<x<-d_F)=\frac{e^{ik'(x+d_F)}+u'e^{-ik'(x+d_F)}}{1+u'}\nonumber\\
    &\times \frac{-\omega\mu_0M_y(Ge^{-ikd_F}+e^{ikd_F})}{
    k(Ge^{ikd_F}-e^{-ikd_F})-k'f(u)(Ge^{ikd_F}+e^{-ikd_F})
    }.
\end{align}
Out of the heterostructure,
\begin{align}
    &E_z(x>d_F+d_S)=\frac{ue^{ik'd_S}+e^{-ik'd_S}}{1+u}\nonumber\\
    &\times \frac{-\omega\mu_0M_y(Ge^{ikd_F}+e^{-ikd_F})e^{ik(x-d_F-d_S)}}{
    k(Ge^{ikd_F}-e^{-ikd_F})-k'f(u)(Ge^{ikd_F}+e^{-ikd_F})
    },\nonumber\end{align}
    \begin{align}
        & E_z(x<-d_F-d_S')=\frac{e^{-ik'd_S'}+u'e^{ik'd_S'}}{1+u'}\nonumber\\
    &\times \frac{-\omega\mu_0M_y(Ge^{-ikd_F}+e^{ikd_F})e^{-ik(x+d_F+d_S)}}{
    k(Ge^{ikd_F}-e^{-ikd_F})-k'f(u)(Ge^{ikd_F}+e^{-ikd_F})
    }. 
    \end{align}

The magnetic field follows $B_y=-\partial_xE_z/(i\omega)$, which inside the magnetic insulator reads  
\begin{align}
   &H_y=-\frac{(2M_yd_F/\lambda) f(u)f(u')}{(f(u)+f(u'))+2 (d_F/\lambda) f(u)f(u')}.
\end{align}
Retaining the leading order in $k$, its real part 
\begin{align}
     \Re(H_y)&\approx- {2 d_FM_y \tanh(d_S/\lambda)\tanh(d_S'/\lambda)}\nonumber\\&\times\left[\lambda(\tanh(d_S/\lambda)+\tanh(d_S'/\lambda))\right.\nonumber\\
     &\left.+2d_F\tanh(d_S/\lambda)\tanh(d_S'/\lambda)\right]^{-1}
\end{align}
leads to the FMR frequency
\begin{align}
    \omega_{\rm K}&=\mu_0\gamma\sqrt{H_0+M_0}
     \sqrt{H_0-M_0\Re(H_y)/M_y},
     \label{general_omega_K}
\end{align}
and its imaginary part
\begin{align}
    \Im({ H}_y)=&2kd_FM_y\left(
    \frac{\tanh^2(d_S'/\lambda)}{\cosh^2(d_S/\lambda)}+\frac{\tanh^2(d_S/\lambda)}{\cosh^2(d_S'/\lambda)}
    \right)\nonumber\\
\times&\left[\tanh(d_S/\lambda)+\tanh(d_S'/\lambda)\right.\nonumber\\+&\left.2d_F/\lambda\tanh(d_S/\lambda)\tanh(d_S'/\lambda)\right]^{-2},
\end{align}
contributes to the damping coefficient 
\begin{align}
    \alpha_R=
    {\mu_0\gamma M_0\Im(H_y)}/({\omega_{\rm K}M_y}).
    \label{general_alpha_R}
\end{align}

\section{Quasi-static approximation in S$|$FI$|$S heterostructure}
 \label{appendix_B}

As justified, the quasi-static approximation $\nabla\times {\bf H}=0$ or ${\bf J}_s$ is allowed when solving the electric fields \textit{near} the heterostructure~\cite{Jackson}.
In the FMR case, the radiated electric field is uniform in the $y$-$z$ plane, so from $\nabla\times {\bf E}=i \omega {\bf B}$, the $x$-component  $B_x=H_{d,x}+M_x=0$ generates no electric field outside the magnet. On the other hand, in the linear response regime for the magnetization dynamics, $M_z=M_0$, so $B_z=\mu_0 (H_0+M_z)$ is static, so only $B_y=\mu_0 M_y$ in the magnet radiates the time-dependent electric field according to
 $-\partial_xE_z=i\omega \mu_0 (M_y+H_{s,y})$.  Integrating along $x$ across the ferromagnet yields the net electric field at the interfaces obeying
\begin{align}
E_z(x=d_F)-E_z(x=-d_F)=-2d_F i \omega \mu_0 (M_y+H_{s,y}).
\label{electric-field}
\end{align}  
Out of the heterostructure, from the $z$-component of $\nabla\times {\bf H}=0$, $H_y|_{\rm outside}$ is a constant, which can be proved to vanish as in Sec.~\ref{quasi_static_bilayer}.

In the quasi-static approximation, the electric field in the superconductors $``1"$ and $``2"$ obeys Eq.~(\ref{electric_field_in_s}). 
From the boundary conditions with  continuous $E_z$
and $H_y$ at interfaces and $H_y|_{\text{outside}}=0$, the electric field in the superconductors reads
 \begin{align}
 &E_z(d_F<x<d_F+d_S)\nonumber\\
 &=E_z(x=d_F)\dfrac{\cosh((x-d_S-d_F)/\lambda)}{\cosh(d_S/\lambda)},\nonumber\\
 &E_z(-d_F-d_S<x<-d_F)\nonumber\\
 &=E_z(x=-d_F)\dfrac{\cosh((x+d_S'+d_F)/\lambda)}{\cosh(d_S'/\lambda)},
 \end{align}  
 which drive the supercurrents in the superconductors adjacent to the magnet.  For thin superconducting films of thickness $O({\lambda})$, we are allowed to take an average of the supercurrents ${\bf J}^{(1)}_s=\left[{\bf J}_s(x=d_F)+{\bf J}_s(x=d_F+d_S)\right]/2 $ and $ {\bf J}^{(2)}_s=\left[{\bf J}_s(x=-d_F)+{\bf J}_s(x=-d_F-d_S)\right]/2 $, i.e.,
\begin{align}
 &{{ J}}_{s,z}^{(1)}=\dfrac{i}{\omega \mu_0\lambda^2}E_z(x=d_F)\dfrac{1+\cosh(d_S/\lambda)}{2\cosh(d_S/\lambda)},\nonumber\\
 &{{ J}}^{(2)}_{s,z}=\dfrac{i}{\omega \mu_0\lambda^2}E_z(x=-d_F)\dfrac{1+\cosh(d_S'/\lambda)}{2\cosh(d_S'/\lambda)}.\label{current}
\end{align}

The supercurrents generate the vector potential (\ref{vector_potential}) and 
the Oersted magnetic field according to ${H}_{s,y}=-\partial_xA_z/\mu_0$.
Using the Weyl identity (\ref{Weyl_identity_2}) we obtain
\begin{align}
	{H_{s,y}}(x)=\left\{
	\begin{array}{cc}
	\left(d_S{{ J}}^{(1)}_{s,z}+d_S'{{ J}}^{(2)}_{s,z}\right)/2,	& x>d_F+d_S \\
	\left(-d_S{{ J}}^{(1)}_{s,z}+d_S'{{ J}}^{(2)}_{s,z}\right)/2,	&~ -d_F<x<d_F \\
\left(-d_S{{ J}}^{(1)}_{s,z}-d_S'{{ J}}^{(2)}_{s,z}\right)/2,	&~~ x<-d_F-d_S'
	\end{array}
	\right. .
\end{align}
 $H_{s,y}|_{\text{outside}}=0 $ requests
\begin{align}
	d_S{{ J}}^{(1)}_{s,z}+d_S'{{ J}}^{(2)}_{s,z}=0
 \label{out_H},
\end{align}  
so the Oersted magnetic field inside the ferromagnetic slab is reduced to 
\begin{align}
{H}_{s,y}(-d_F<x<d_F)=d_S'{J}_{s,z}^{(2)}=-d_SJ_{s,z}^{(1)}.\label{in_H}
\end{align} 
Thereby, when $d_S=d_S'$, the supercurrents are opposite in the two superconductors.
When $d_S'\rightarrow 0$, $H_{s,y}$ vanishes in the magnet.

Substituting Eqs.~(\ref{current}) and (\ref{electric-field}) into (\ref{out_H}), we obtain the electric field at the surface of the ferromagnetic film:
   \begin{align}
   \nonumber
&E_z(x=-d_F)=i \mu_0\omega d_Sd_F (M_y+H_{s,y}) \dfrac{\cosh(d_S/\lambda)+1}
{\cosh(d_S/\lambda)}\\
&\times \left(
\dfrac{d_S(\cosh(d_S/\lambda)+1)}{2\cosh(d_S/\lambda)}+\dfrac{d_S'(\cosh(d_S'/\lambda)+1)}{2\cosh(d_S'/\lambda)}
\right)^{-1}.
\end{align}  
Substituting it into Eq.~(\ref{in_H}), the Oersted magnetic field in the ferromagnetic film
\begin{align}
{H}_{s,y}(-d_F<x<d_F)&= -M_y\dfrac{ d_Fd_S'd_S G(d_S,d_S',\lambda)}{\lambda^2+d_Fd_S'd_S G(d_S,d_S',\lambda)}  , 
\label{H_y}                     
\end{align} 
where 
    \begin{align}
    \nonumber
&G(d_S,d_S',\lambda) =\dfrac{(\cosh(d_S/\lambda)+1)}{\cosh(d_S/\lambda)}\dfrac{(\cosh(d_S'/\lambda)+1)}{\cosh(d_S'/\lambda)}\\
&\times\left(
\dfrac{d_S(\cosh(d_S/\lambda)+1)}{\cosh(d_S/\lambda)}+\dfrac{d_S'(\cosh(d_S'/\lambda)+1)}{\cosh(d_S'/\lambda)}\right)^{-1}.
\end{align} 
These results capture precisely the key physics of the full solution and are convenient for the calculation of the interaction between Kittel magnon and Cooper-pair supercurrent.

In the linear regime of the magnetization dynamics, substituting $B_x=M_x+H_{d,x}=0$ into the Landau-Lifshitz equation
\begin{align}
    &-i\omega M_x+\mu_0\gamma M_yH_0=\mu_0\gamma M_0H_{s,y},\nonumber\\
    &i\omega M_y+\mu_0\gamma M_x H_0=\mu_0\gamma M_0H_{d,x},
    \label{LLG-equation}
\end{align}
we find $M_y$ relates to $H_{s,y}$ via 
\begin{align}
M_y=\dfrac{\mu_0^2\gamma^2M_0(H_0+M_0)}{\mu_0^2\gamma^2H_0(H_0+M_0)-\omega^2}H_{s,y}.\label{M_y}
\end{align}
    When $d_S'\rightarrow0$, $H_{s,y}=0$ according to Eq.~(\ref{H_y}), and the FMR frequency recovers the Kittel formula $\tilde{\omega}_{\rm K}=\mu_0\gamma\sqrt{H_0(H_0+M_0)}$~\cite{kittel_mode}. With finite $d_S$ and $d_S'$, the FMR frequency is self-consistently solved via combining Eqs.~(\ref{H_y}) and (\ref{M_y}), leading to the modified FMR frequency 
\begin{align}
	&\omega_{\rm K}=\mu_0 \gamma\nonumber
 \\ &\times\sqrt{\frac{\lambda^2 H_0(H_0+M_0)+d_Sd_S'd_FG(d_S,d_S',\lambda)(H_0+M_0)^2}{d_Sd_S'd_FG(d_S,d_S',\lambda)+\lambda^2}}.
\end{align}
In particular, when $ d_S=d_S' $, 
  \begin{align}
\omega_{\rm K}&= {\mu_0 \gamma}\left(\dfrac{2\lambda^2 \cosh{(d_S/\lambda)}H_0(H_0+M_0)}
{d_Sd_F\left(\cosh{(d_S/\lambda)}+1\right)+2\lambda^2\cosh{(d_S/\lambda)}}\right.\nonumber\\
&\left.+\dfrac{d_Sd_F(\cosh{(d_S/\lambda)}+1)(H_0+M_0)^2}{d_Sd_F\left(\cosh{(d_S/\lambda)}+1\right)+2\lambda^2\cosh{(d_S/\lambda)}}\right)^{1/2}.
\label{FMR_shifted}
\end{align}  
Approaching $T_c$, $\lambda\rightarrow \infty$, $\cosh(d_S/\lambda)\rightarrow 1$, so the FMR frequency (\ref{FMR_shifted}) recovers the Kittel formula $\omega_{\rm K}\rightarrow \tilde{\omega}_{\rm K}$; otherwise $T<T_c$, it is shifted.

\end{appendix}

\end{document}